\documentclass[fleqn,usenatbib]{mnras}

\usepackage{newtxtext,newtxmath}

\usepackage[T1]{fontenc}

\DeclareRobustCommand{\VAN}[3]{#2}
\let\VANthebibliography\thebibliography
\def\thebibliography{\DeclareRobustCommand{\VAN}[3]{##3}\VANthebibliography}

\usepackage{graphicx}	
\usepackage{amsmath}	
\usepackage{orcidlink}

\title[Ca-bearing cyanopolyynes in IRC+10216]{Calcium-bearing cyanopolyynes in IRC+10216}

\author[T. J. Millar]{
T. J. Millar\orcidlink{0000-0001-5178-3656}\thanks{E-mail: tom.millar@qub.ac.uk}
\\
\\
Astrophysics Research Centre, School of Mathematics and Physics, Queen's University Belfast, University Road, Belfast BT7 1NN, UK\\
}

\date{Accepted 2026 May 25. Received 2026 May 15; in original form 2026 April 09}

\pubyear{\the\year{}}

\begin{document}
\label{firstpage}
\pagerange{\pageref{firstpage}--\pageref{lastpage}}
\maketitle

\begin{abstract}
In recent years, a number of metal-containing, carbon-chain species have been detected 
in the external circumstellar envelope of the carbon-rich AGB star IRC+10216.  
The most common metal detected in such species is Mg, for which molecules as large as 
MgC$_5$N, MgC$_5$N$^+$, MgC$_6$H and MgC$_6$H$^+$ have been observed.  In this paper, 
we calculate the likely abundances of the Ca-bearing cyanopolyynes, CaC$_{2n+1}$N 
for n = 1-4, drawing the conclusion that the observed abundance of CaNC must be 
made from much larger Ca-terminated cyanopolyyne ions, which requires considerable 
rearrangement in their dissociative recombination. We pay particular attention to the 
detectability of CaC$_3$N whose rotational spectrum has recently been measured.

\end{abstract}




\section{Introduction}
\label{sec:intro}

The presence of long carbon-chain molecules, such as the cyanopolyynes HC$_{2n+1}$N, n $=$ 1--4,
in the circumstellar envelope (CSE) of the carbon-rich Asymptotic Giant Branch (AGB) star 
IRC+10216  has been known for many years.  
Their formation is driven by interstellar ultraviolet photons which penetrate the CSE, 
dissociating and ionising stable molecules, such as HCN and $\rm{C_2H_2}$, thereby triggering 
a complex network of neutral-neutral and ion-neutral reactions which build these long chain 
species \citep{mil94, mil00, mil24}.

Using optical spectroscopy of a background star in the outer CSE of IRC+10216, \citet{mau10}  
detected absorption features due to atomic metals, including Na, K, Ca, Cr and Fe, thus 
showing that these elements were not completed depleted by dust condensation in the inner 
CSE. 
Subsequently, there has been the detection of rotational emission lines from metal-bearing, 
long chain polyacetylides, XC$_{2n}$H, n = 1--3, and cyanopolyynes, XC$_{2n+1}$N, 
n = 1--2, \citep{agu14, cer19, par21} for X = Mg, as well as the ions MgC$_2$H$^+$, MgC$_4$H$^+$, 
MgC$_3$N$^+$, and MgC$_5$N$^+$ \citep{cer23}.  
The surprising detection of these ions in the CSE is helped by the fact that their reactions 
with H$_2$ are calculated to be endothermic \citep{cer23}. Other Mg-bearing molecules 
detected include HMgNC \citep{cab13}, HMgC$_3$N \citep{cab23} and MgC$_2$ \citep{cha22}.
Related detections include NaC$_3$N \citep{cab23} and the linear silicon carbides, 
SiC$_4$, SiC$_5$ and SiC$_6$ \citep{par25, cer25}.  
Observations of these species are consistent with their formation in the outer CSE, 
that is, in the same region of the envelope in which the metal atoms are observed 
\citep{mau10}.

Motivated by these detections and those of CaNC \citep{cer19b} and CaC$_2$ \citep{gup24}, 
we present in this paper the first calculation that considers the formation of 
calcium-bearing cyanopolyymes and estimate their column densities in a photochemical 
kinetic model for the outer CSE of IRC+10216.

\section{Chemical Model}
\label{sec:model}

Chemical kinetic schemes for the formation of metallic polyacetylides and cyanopolyynes  
are almost entirely based on the pioneering work of Petrie and co-workers 
\citep{pet96, pet97, pet00, dun02, pet04} who suggested that the metal-terminated 
cyanopolyynes were formed through the radiative association of metal atomic ions 
with the cyanopolyynes, for example: 

\begin{equation}
\rm{Mg^+ + HC_5N  \rightarrow  
 MgC_5NH^+ + h\nu}   
\end{equation}
followed by dissociative recombination with electrons to give products such as:

\begin{equation}
\rm{MgC_5NH^+ + e^-  \rightarrow  MgC_5N + H} 
\end{equation}
and possibly other channels, noting that rate coefficients,  products, and branching 
ratios are unknown.

\citet{cer23} and \citet{mil24} have fitted  the theoretical calculations of Petrie and 
colleagues to the typical de Kooij-Arrhenius formula widely used in interstellar chemical 
models and presented the results of the syntheses of Mg-bearing molecules in IRC+10216.  
Two general inferences can be drawn from these.  One is that the radiative associations 
between metal ions and HCN and HNC are very -- indeed vanishingly -- small, but increase 
by several orders of magnitude, and approach the collisional rate at low temperatures, as 
the size of the neutral molecule increases. The second is that due to this, the column 
densities of the smallest, triatomic species, such as MgCN and MgNC, which tend to be the 
largest, are severely underproduced if they are form only from their protonated ions.  
As a result, it is necessary that the dissociative recombination of the larger ions must 
also produce the abundant triatomic molecules. For this reason, \citet{par21}, considering 
the dissociative recombination of MgC$_5$NH$^+$, adopted branching ratios of 0.74, 0.25 
and 0.01, forming MgNC, MgC$_5$N and HMgNC, respectively.

To date, only two Ca-bearing molecules have been detected in the outer CSE of IRC+10216: 
CaNC \citep{cer19b} with a column density of 2 $\times$ 10$^{11}$ cm$^{-2}$, and CaC$_2$ 
with a column density of 5 $\times$ 10$^{11}$ cm$^{-2}$  \citep{gup24}. The formation of 
the latter molecule and the CaC$_{2n}$H, n $\ge$ 1, species is unknown, although 
\citet{koc25} have studied the reaction of ground state Ca$^+$ 
with C$_2$H$_2$ at very low temperatures and find that it is unreactive to form 
CaC$_2$H$^+$ and, further, that the radiative association to form CaC$_2$H$_2^+$ is 
endothermic.

\subsection{Radiative Association}
\label{sec:rad}

\citet{pet04} made a theoretical study of the radiative association reactions of Ca$^+$ 
with HCN and the cyanopolyynes HC$_3$N, HC$_5$N and HC$_7$N, determining bond energies, 
structures, and the rate coefficients at 30 K. 
Since Petrie (2004) calculated rate coefficients only at 30 K, we adopt the same temperature 
dependence as those of the corresponding Mg$^+$ reactions. In particular, we fit 
the values calculated by \citet{dun02} over 10--300 K to the form:

  \begin{equation}
  \mathrm{k(T)} = \alpha (\mathrm{T}/300)^\beta \quad \mathrm{cm^3 s^{-1}} \,,
  \end{equation} 
to derive the $\beta$ values and subsequently determine $\alpha$ by fitting to 
Petrie's values at 30\,K.

Petrie comments that his 30 K value 
for HC$_7$N is non-physical due to the neglect of collisional saturation. 
Instead, for this and for HC$_9$N, we calculate their values using the collisional rate
 formula \citep{su82}:
   \begin{equation}
    \mathrm{k(T)} = 3.87 \times 10^{-9} (\mu_\mathrm{D}/\mu)^{1/2} (\mathrm{T}/300)^{-1/2} \quad \mathrm{cm^3 s^{-1}} \,, 
    \end{equation}
 where $\mu_D$ is the electric dipole moment of the neutral molecule in Debye and $\mu$ is 
 the reduced mass of the reactants in atomic mass units. In this limit, every collision 
 leads to radiative association.

 Table \ref{tab:cacyano} gives the data used to calculate the radiative association reaction 
 rate coefficients as a function of temperature.

 \begin{table}
 \centering
 \caption{Radiative association rate coefficients of Ca$^+$ with cyanopolyynes. 
 Here, $\alpha$ is the value at 300 K and $\beta$ is the temperature dependence. }  
 \label{tab:cacyano}
\begin{tabular}{lll}
\hline 
Neutral & $\alpha$ & $\beta$ \\
\hline
HCN & 7.09 $\times$ 10$^{-18}$  & -1.49 \\
HC$_3$N & 1.31 $\times$ 10$^{-14}$ & -1.65  \\
HC$_5$N & 6.99 $\times$ 10$^{-11}$ & -1.32 \\
HC$_7$N & 3.49 $\times$ 10$^{-9}$ & -0.50  \\
HC$_9$N & 3.66 $\times$ 10$^{-9}$ & -0.50  \\
\hline
\end{tabular}
\end{table}

In general, \citet{pet04} finds that the most stable products are ions of the form 
CaNC$_{2n+1}$H$^+$ rather than CaC$_{2n+1}$NH$^+$ and we incorporate this in our 
reaction scheme.

\subsection{Ion-neutral and Photoreactions}
\label{sec:ion}

The Ca-bearing cyanopolyynes have large electric dipole moments and proton affinities 
(Table~\ref{tab:cadata}).  As a result, they have very fast reactions with ions, 
most importantly proton transfer with HCO$^+$ and bond breaking via C$^+$ in the CSE. 
In the latter case, we assume products are Ca$^+$, C and C$_{2n+1}$N.  This maximises 
destruction rates by ensuring that there is no efficient recycling of the parent neutral.
We have calculated all ion-neutral rate 
coefficients using the Su-Chesnovich formula.  
Due to a lack of thermodynamic data and theoretical and experimental kinetic information, 
we have not considered reactions involving
neutral species. Collisions with C atoms and C$_2$H radicals might play a role, the former in 
the loss of Ca-cyanopolyynes, the latter in either loss or growth of their carbon backbone.
We discuss this further in Sec.~\ref{sec:results}.

\begin{table}
 \centering
 \caption{The proton affinities (PA) in kJ mol$^{-1}$ and the dipole moment, $\mu_D$ in 
 Debye for CaC$_5$N are taken from \citet{pet04}.  The dipole moments fo CaNC and CaC$_3$N 
 are from the CDMS database \citep{mul05} and \citet{gur25}, respectively. Those for the 
 larger molecules are extrapolated from the smaller species. }  
 \label{tab:cadata}
\begin{tabular}{lll}
\hline 
Molecule & $\mu_D$ & PA \\
\hline
CaNC & 6.90 & 961.9 \\
CaC$_3$N & 7.87 & 859.5  \\
CaC$_5$N & 9.81 & 846.8  \\
CaC$_7$N & 10.80 &  - \\
CaC$_9$N & 11.80 &  - \\
\hline
\end{tabular}
\end{table}

The photodissociation rates and products of the Ca-cyanopolyynes are unknown, so we adopt 
an approach that minimises the possibility of their products to easily recycle to the 
original neutral. Since the Ca-ligand bond is typically the weakest in the neutral, we 
simply assume that the only product channel is:

\begin{equation}
  \rm{CaC_{2n+1}N + h\nu \rightarrow Ca + C_{2n+1}}N.      
\end{equation}

For consistency with the discussion of the Mg-cyanopolyynes in \citet{mil24}, we adopt 
the same photodissociation rate for all, 10$^{-10}$\textrm{e}$^{-1.7{A_V}}$ \textrm{s}$^{-1}$. 
We account for the fact that the circumstellar envelope itself provides shielding over 4$\pi$ 
steradians against the interstellar radiation field \citep{jur81}. We discuss the 
effects of choosing larger and smaller unshielded photorates on abundances in Sec.~\ref{sec:results}.

\subsection{Dissociative Recombination}
\label{sec:dissrec}

Dissociative recombination of the protonated species formed in radiative association is 
likely to be a fast process, although the neutral products are very uncertain, with multiple 
products possible for the largest ions.  As noted earlier in Sec.~\ref{sec:model}, the observed 
abundances of metal cyanides and isocyanides in IRC+10216, coupled with very low values 
of the relevant radiative association rates, mean that dissociative recombination of the 
larger radiative association products must also form the triatomic molecules.  We provide 
results for three possible schemes to illustrate their influence on radial abundance profiles 
and radial column densities. 

One, Model A, asumes that the ions CaNC$_{2n+1}$H$^+$ recombine only to the abundant CaNC 
molecule and to CaC$_{2n+1}$N in the ratio 3:1. This minimises the contribution of 
larger ions to the formation of smaller Ca-cyanopolyynes.  Thus, for CaNC$_3$H$^+$:

\begin{eqnarray}
  \rm{CaNC_3H^+ + e^-} & \rightarrow & \rm{CaNC + C_2H}  \\
  \rm{CaNC_3H^+ + e^- } & \rightarrow & \rm{CaC_3N + H}   
\end{eqnarray}

A second approach, Model B, modifies this
by setting the branching ratios for CaNC$_5$H$^+$ to be similar to those used by 
\citet{par21} for MgC$_5$NH$^+$ in which these were determined by fitting the observed 
abundances of four Mg-bearing molecules relative to that of MgNC. Ignoring channels with
branching ratios less than 10 percent, which also form species whose Ca equivalents 
are not in  our model, we find the neutral products are CaNC, CaC$_3$N and CaC$_5$N, with 
branching ratios 1/2, 1/3 and 1/6, respectively.

The final approach, Model C, follows the prescription from \citet{pet04} 
who discussed the energetics 
involved in the dissociative recombination of CaNCH$^+$, CaNC$_3$H$^+$ and CaNC$_5$H$^+$. 
He noted that his results could be described by a few general rules: (i) triple products 
are all likely to be endoergic and do not proceed, (ii) 75\% of reactions take place on 
triplet surfaces,  (iii) it is very unlikely that triple C--C or C--N bonds break, and 
(iv) products such as CaCN, CaC$_3$N, CaC$_5$N, etc. are not directly accessible in the 
dissociative recombination process. He did, however, suggest that the CaC$_{2n+1}$N 
molecules are plausible products due to low barriers to isomerisation of the CaNC$_{2n+1}$ 
species during the process of dissociative recombination. 

It is noteworthy that \citet{dun02} showed that, for M = Na, Mg and Al, 
all the M$^+$--HC$_{2n+1}$N adducts formed by radiative association have the structure 
MNC$_{2n+1}$H$^+$. The observations of the Mg-bearing cyanopolyynes and NaC$_3$N in IRC+10216
tend to the conclusion that isomerisation is common in the recombination process although
no direct proof has yet been shown.

As an illustration of our implementation, consider the recombination of CaNC$_3$H$^+$ on 
the singlet surface with overall branching ratio 1/4. Petrie shows that there are three 
exothermic channels available, producing Ca, CaNC and CaC$_3$N.
We do not know the branching ratios to each of these, so we make the assumption that all 
three are equal. Thus, each of these products accounts for 1/12th of the total rate 
constant. For reaction on the triplet surface, with branching ratio 3/4, there are only 
two products allowed, CaNC and CaC$_3$N. Again assuming equal branching ratios, 
they each account for 3/8ths of the triplet surface products.  The total branching ratios 
are given by the sum of the singlet and triplet values, 1/12, 11/24 and 11/24 for Ca, CaNC 
and CaC$_3$N, respectively. We adopt a total rate coefficient of 3.0 $\times$ 
10$^{-7}$ (T/300)$^{-1/2}$ cm$^3$ s$^{-1}$ for each ion. 

\section{Model Description}
\label{sec:modesc}

We have included the new calcium species and their relevant reactions in an extension of 
the RATE22 version\footnote{https://www.umistdatabase.uk} of the UMIST Database for 
Astrochemistry (UDfA) \citep{mil24}. This release included only two Ca-bearing molecules, 
CaO and CaOH, but their description is restricted to high temperature reactions and they 
play no part in the carbon-rich environment of IRC+10216.  

Our physical model for the expanding CSE gas is the same as that described in 
\citet{mil24} -- a uniform mass-loss rate of 3.0 $\times$ 10$^{-5}$ M$_{\odot}$ yr$^{-1}$, 
an expansion velocity of 14.5 km s$^{-1}$, and a power-law temperature distribution, 
T(r) = T$_*$(r/R$_*$)$^{-0.7}$, where the stellar temperature and radius are 2330~K and 
5 $\times$ 10$^{13}$ cm, respectively.
\citet{mau10} detected absorption lines from both \ion{Ca}{i} and \ion{Ca}{ii} in the outer 
CSE of IRC+10216 at a projected distance of 35 arcsec, deducing column densities of 
1.9 $\times$ 10$^{12}$ and 7.0 $\times$ 10$^{12}$ cm$^{-2}$, respectively. We calculate 
abundances from an inner radius r = 10$^{15}$ cm, where T = 286 K and n(H$_2$) = 2.33 $\times$ 10$^7$ cm$^{-3}$, 
to 10$^{18}$ cm$^{-2}$, where T = 10 K and n(H$_2$) = 23.3 cm$^{-3}$. At our initial radius, 
the radial visual extinction is 25 mag. For a smooth, constant velocity outflow, the visual 
extinction in the radial direction is proportional to 1/r.
 Finally, we take the initial abundance of atomic calcium to be 7.0 $\times$ 10$^{-7}$ 
 relative to H$_2$, that is, depleted by a factor of about  six relative to its 
 solar value. This value is somewhat arbitrary but leads to a CaNC column 
 density in all three models similar to that observed.

 \section{Model Results}
 \label{sec:results}

 Table~\ref{tab:coldens} provides the calculated radial column densities of Ca-bearing 
 cyanopolyynes for branching ratios following the recipes proposed in Sec.~\ref{sec:dissrec},  
 by limiting the products to CaNC and one Ca-bearing cyanopolyyne (Model A), subsequently adding  the 
 branching ratios used by \citet{par21} for CaNC$_5$H$^+$ (Model B), and finally those that 
 consider the multiple product channels proposed by
 \citet{pet04} (Model C). 
 
\begin{figure}
	\includegraphics[width=\columnwidth]{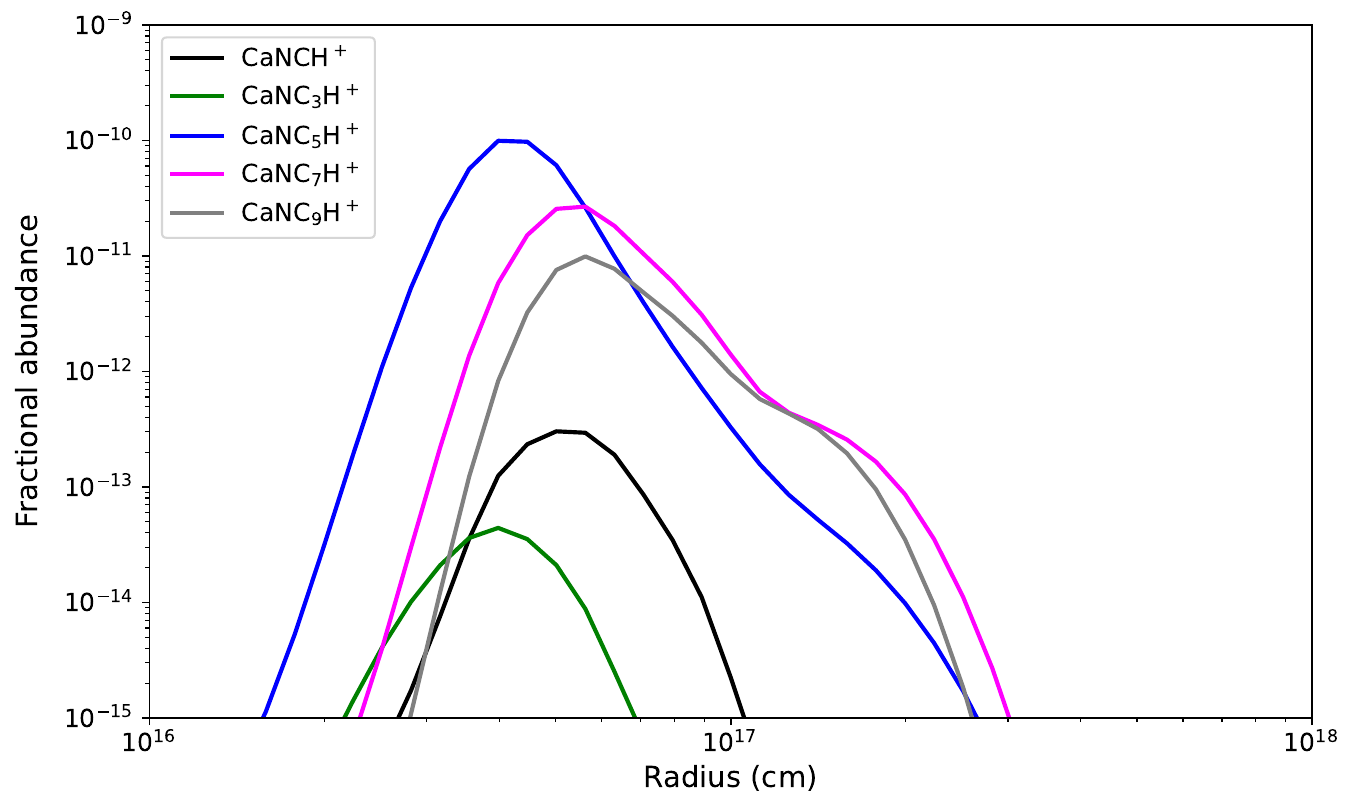}
    \caption{Fractional abundances relative to H$_2$ of the protonated Ca-bearing 
    cyanopolyynes as a function of radius in Model A.}
    \label{fig:modelA_ions}
\end{figure}

The major difference between models occurs for CaC$_3$N which, in Model A, has a column density about 
 1000 times less than that in Models B and C.  This is a consequence of the fact that in Model A,
 it is formed only through recombination of CaNC$_3$H$^+$, formed through a very slow 
 radiative association between Ca$^+$ and HC$_3$N (see Table~\ref{tab:cacyano}).
 It has the lowest abundance of all the protonated Ca-cyanopolyynes in every model -- 
 see Fig.~\ref{fig:modelA_ions} for the protonated ion abundances in Model A; other models 
 have very similar distributions.

 In Model B, however, we allow CaC$_3$N to form from the recombination of CaNC$_5$H$^+$, 
 the most abundant ion, with the result that its column density increases dramatically.

 Finally, Model C, in which all recombinations lead to multiple neutrals, shows 
 further increases in column density, with the exception of CaC$_9$N, which forms only from 
 CaNC$_9$H$^+$ and with a smaller branching ratio in Model C than in Models A and B.

\begin{table}
 \centering
 \caption{Calculated radial column densities (cm$^{-2}$).} 
 \label{tab:coldens}
\begin{tabular}{llll}
\hline 
Molecule & Model A & Model B & Model C\\
\hline
CaNC & 6.87 $\times$ 10$^{11}$ & 5.24 $\times$ 10$^{11}$ & 2.65 $\times$ 10$^{11}$ \\
CaC$_3$N & 1.88 $\times$ 10$^{8}$ & 2.13 $\times$ 10$^{11}$ & 2.56 $\times$ 10$^{11}$ \\
CaC$_5$N & 3.64 $\times$ 10$^{11}$ & 3.12 $\times$ 10$^{11}$ & 4.56 $\times$ 10$^{11}$ \\
CaC$_7$N & 9.53 $\times$ 10$^{10}$ & 9.53 $\times$ 10$^{10}$ & 1.05 $\times$ 10$^{11}$ \\
CaC$_9$N & 3.15 $\times$ 10$^{10}$ & 3.15 $\times$ 10$^{10}$ & 2.81 $\times$ 10$^{10}$ \\
\hline
\end{tabular}
\end{table}

 The column density of CaNC decreases from Model A to Model C as the branching ratios
 for its formation decrease due the increasing number of product channels
 from Model A to B to C. 
 
In our models, the only Ca-bearing molecule observed in IRC+10216 is CaNC with column 
density, N = 2.0 $\times$ 10$^{11}$ cm$^{-2}$ \citep{cer19b}. All models reproduce 
the observed value reasonably well, given the many assumptions made here.
 
\begin{figure}
	\includegraphics[width=\columnwidth]{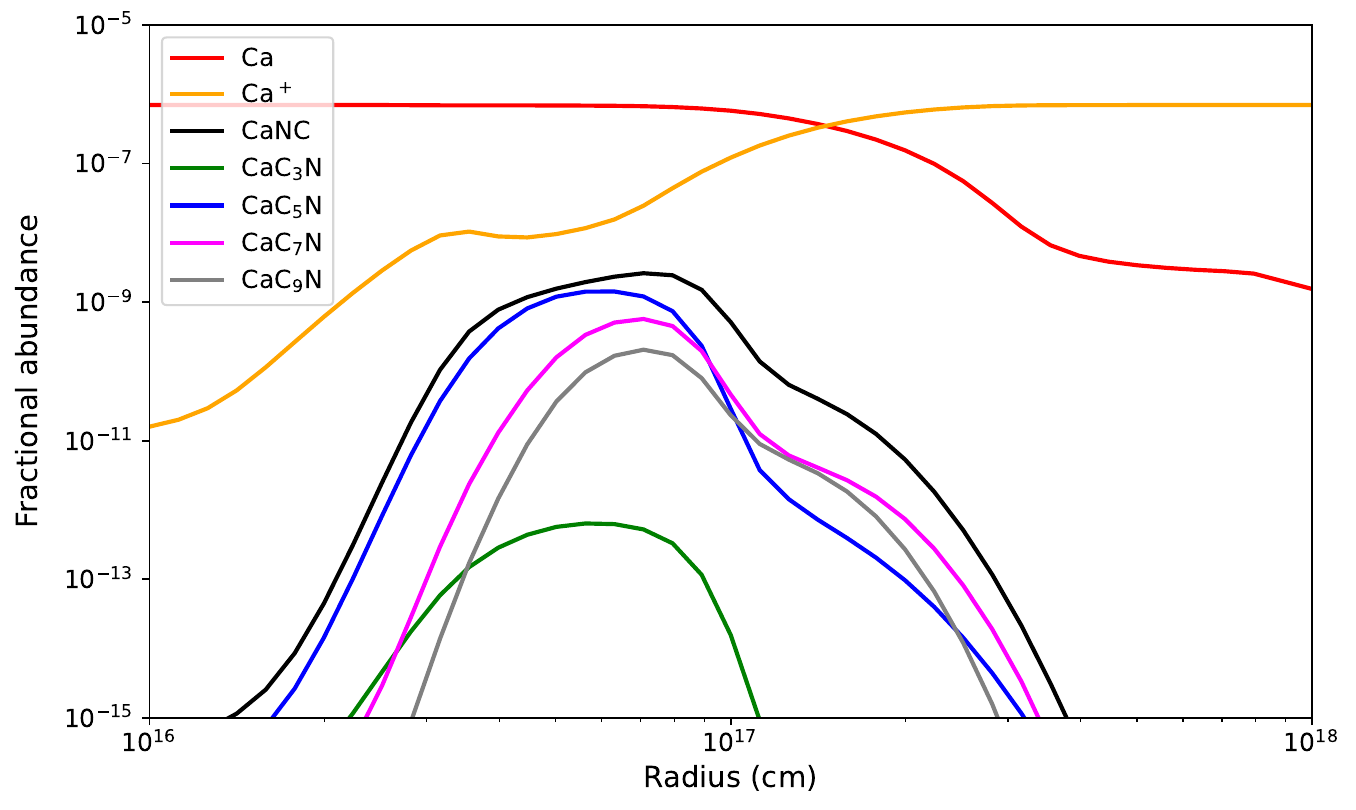}
    \caption{Fractional abundances relative to H$_2$ of the Ca-bearing cyanopolyynes 
    as a function of radius in Model A.}
    \label{fig:modelA}
\end{figure}

\begin{figure}
	\includegraphics[width=\columnwidth]{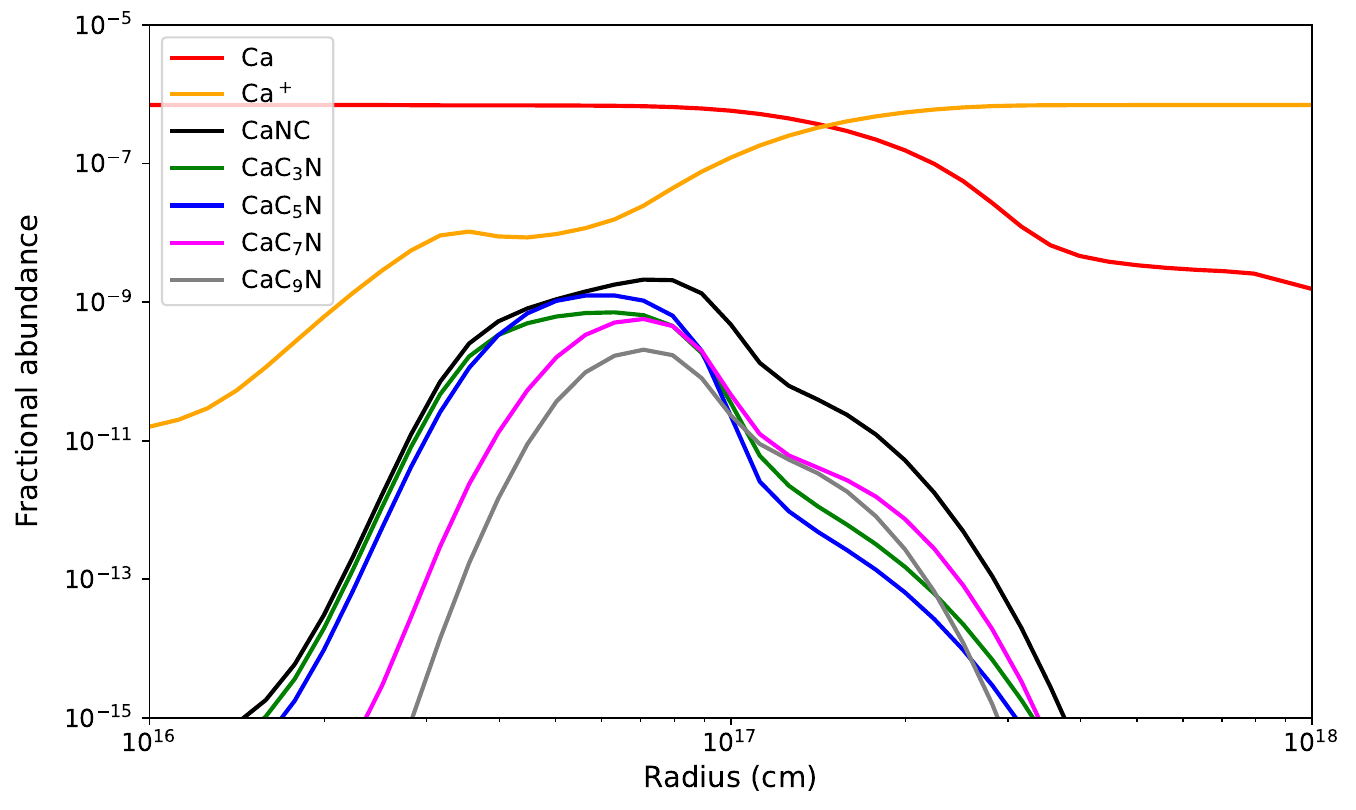}
    \caption{Fractional abundances relative to H$_2$ of the Ca-bearing cyanopolyynes 
    as a function of radius in Model B.}
    \label{fig:modelB}
    \end{figure}

\begin{figure}
	\includegraphics[width=\columnwidth]{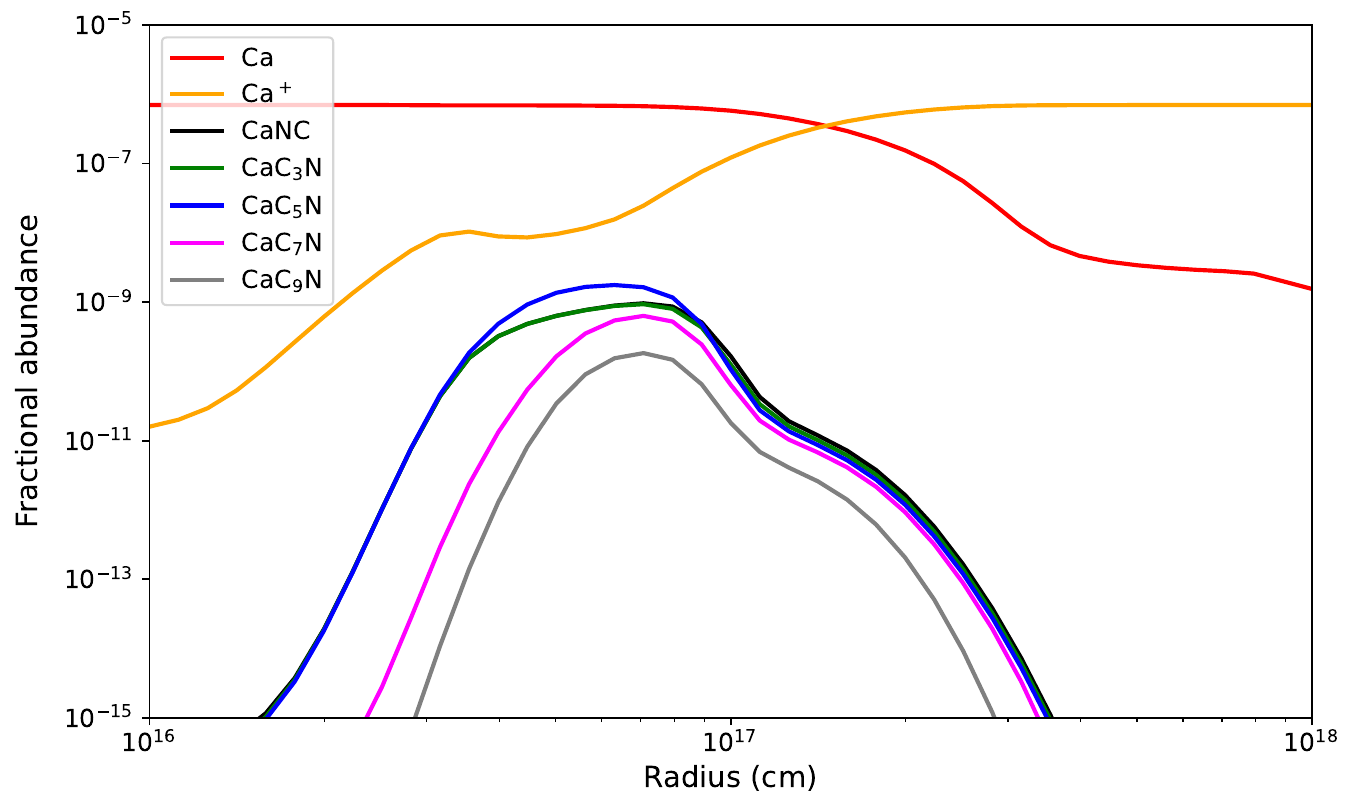}
    \caption{Fractional abundances relative to H$_2$ of the Ca-bearing cyanopolyynes 
    as a function of radius in Model C.}
    \label{fig:modelC}
\end{figure}

 Figs.~\ref{fig:modelA}, \ref{fig:modelB} and \ref{fig:modelC} show the fractional 
 abundances  of the Ca-bearing molecules as a function of radius in Models A, B and C, 
 respectively.
 The only significant differences are the much lower abundance of 
 CaC$_3$N in Model B, as discussed above, and the reduction in CaNC fractional abundances 
 and column density in Model C. 
 This results from the reduction of its product branching ratios
 as the size of the ion increases, 
 from 3/4 for all ions in Model A, to 11/24 for the recombination of CaNC$_5$H$^+$ down to 23/120 for
 CaNC$_9$H$^+$ in Model C.

  We have also calculated models in which we varied the unshielded photorates of the 
 Ca-bearing molecules, increasing them by a factor of 3 and lowering them by 10. 
 Neither of these has much effect on column densities, with differences on the order of 
 10--20 percent.  The reason is that, at the radius at which peak fractional abundances 
 are found, destruction is dominated by abundant C$^+$ ions formed through the 
 photodissociation of abundant `parent' molecules such as C$_2$H$_2$ and HCN. The C$^+$ 
 reactions have very fast rate coefficients -- that with CaC$_5$N has a value of 
 3.64 $\times$ 10$^{-8}$ cm$^3$ s$^{-1}$ at 30\,K -- due to the large dipole moments 
 of the neutral molecules.

We have not included neutral-neutral reactions in our chemical model but it
 is possible that they could also participate, although quantification is
 difficult due to the lack of theoretical or experimental information.  For example, in
 analogy with the cyanopolyynes, chain length could increase through reaction 
 with C$_2$H, although this is a destruction mechanism for the smaller 
 Ca-chains.  In addition, the formation of small species that fuel growth, such as 
 CaNC and CaC$_3$N, depends on the dissociative recombination of the larger protonated
 Ca-bearing cyanopolyynes. A further destruction mechanism for the chains could be 
reaction with C atoms, the most abundant atoms in the region of interest. We 
estimate, however, that rate coefficients in excess of 10$^{-10}$ cm$^3$ s$^{-1}$ 
are needed to dominate the very fast ion-neutral reactions that destroy the 
chains.

\subsection{Observability of CaC$_3$N}
\label{sec:obs}

Given that \citet{gur25} have measured the rotational spectrum of CaC$_3$N, as well as 
CaC$_4$H,  we provide an estimate of its rotational line intensities relative to those of 
MgC$_3$N.  Neglecting the unknown effects of infrared pumping in their excitation, 
noting that both have $^2\Sigma^+$ ground states and that both should be present in the same region of the CSE,
we can write the ratio of line intensities as:

\begin{equation}
 \mathrm
 {
    \frac{I_{Ca}}{I_{Mg}} = \left(\frac{\mu_D(Ca)}{\mu_D(Mg)}\right)^2  \frac{B(Ca)} {B(Mg)} \frac{N(Ca)}{N(Mg)}
  }  
\end{equation}
where Ca and Mg refer to CaC$_3$N and MgC$_3$N, respectively, N is column density
and the dipole moment of MgC$_3$N is 6.38 D from the CDMS database 
\citep{mul05}.  We used values for the rotational constants, B(Ca) and B(Mg), of 979.08 and 
1380.89 MHz for CaC$_3$N and MgC$_3$N, from \citet{gur25} and \citet{cer19}, respectively. 
Since the rotational constants are small, k$_{\rm{B}}$T $>$ hB, we use the 
high-temperature approximation for the rotational partition functions. Taking the column density 
calculated in Model C for CaC$_3$N and the observed value for MgC$_3$N of 
9.3 $\times$ 10$^{12}$ cm$^{-2}$ from \citet{cer19}, we find that the line intensity 
ratio is 0.03. The MgC$_3$N lines observed by \citet{cer19} have typical intensities 
of 5\,mK, so that CaC$_3$N is unlikely to be detectable in IRC+10216 without considerable 
effort.

What is an optimistic ratio? We note that as a minor element whose chemistry does not 
interact widely with other species, the fractional abundances, column densities, and line 
intensities of the Ca-bearing species are directly proportional to the initial adopted Ca 
abundance.   Thus, we could increase the Ca abundance to its solar value, a 
factor of about 6. We could also adjust the branching ratios in dissociative recombination.
The most abundant ion is CaNC$_5$H$^+$ with a branching ratio of 5/16 to form CaC$_3$N (Model A). 
Since these larger ions must also form CaNC, the maximum branching ratio allowable is 
perhaps 1/2, an increase of 1.6. In conclusion, we can imagine a situation, though perhaps 
unlikely, in which the intensity ratio increases by an order of magnitude to give lines at 
the 1.5\,mK level.

\section{Conclusions}

We have used theoretical calculations of the radiative association rate coefficients 
between Ca$^+$ and the cyanopolyyens \citep{pet04} in an extended version of the RATE22 
release of the UMIST Database for Astrochemistry. 
We have estimated the abundances of the Ca-bearing cyanopolyynes in the outer envelope of 
IRC+10216, the most likely source of such species given the presence there of several 
Mg-bearing cyanopolyynes and polyacetylides.  
Our study shows that the ions formed in such reactions must undergo significant 
rearrangement, from the lowest energy ions, CaNC$_{2n+1}$H$^+$ to the lowest energy 
neutrals, CaC$_{2n+1}$N, during dissociative recombination with electrons. 

Since the radiative association reactions of Ca$^+$ with both HCN and HC$_3$N are 
very inefficient, even at the low temperature of the outer CSE, the observed abundance 
of CaNC, and a potentially observable abundance of CaC$_3$N, require that dissociative 
recombination of the larger ions must also form these smaller neutral species.  
In particular, the calculated column density of CaC$_3$N is about 1000 times 
higher in Models B and C, in which all larger Ca-bearing ions contribute to its 
formation, than in Model A where it only forms from CaNC$_3$H$^+$.  
All models reproduce the observed CaNC abundance reasonably well.

We also investigated models with larger and smaller photodissociation rates but find that 
they make only small differences to the calculated column densities. This is because in the 
region where fractional abundances peak, their loss is dominated by reaction with abundant 
C$^+$ ions. 

Although there are many uncertainties in the calculations, in particular, the adopted 
calcium abundance and the branching ratios of dissociative recombination, the detection 
of CaC$_3$N in IRC+10216 using the recent measurement of its rotational spectrum \citep{gur25} 
will require considerably more sensitive observations than those for MgC$_3$N.

Finally, we note that more accurate model calculation of the Ca-bearing 
cyanopolyyne column densities will require more modern theoretical and experimental 
treatments of the relevant radiative association and dissociative recombination
reactions.

\section*{Acknowledgements}

I thank the anonymous reviewer for a careful, thorough and helpful report which 
has improved this manuscript. I am grateful to Elvire de Beck and her colleagues at the 
Department of Earth, Space and Environment, Chalmers University of Technology, for 
hospitality during a visit at which this work was conceived. TJM's research at Queen's 
University Belfast is supported by grant ST/T000198/1 from the STFC.

\section*{Data Availability}

 Details of the additional reactions related to the calcium chemistry described here can 
 be obtained upon request to the author.



\bibliographystyle{mnras}
\bibliography{calcium} 



\bsp	
\label{lastpage}
\end{document}